\documentclass[prb,twocolumn,balancelastpage,floatfix,showpacs,letterpaper]{revtex4-1}
\usepackage{times}
\usepackage{verbatim}
\usepackage{bm}
\usepackage{bbm}
\usepackage{pifont}
\usepackage{amsmath}
\usepackage{amsfonts}
\usepackage{amscd}
\usepackage{amsthm}
\usepackage{amssymb}
\usepackage{graphicx}
\graphicspath{{../figures/}{/}}
\usepackage{hyperref}

\newcommand{\ee}                        {e}                                  
\newcommand{\idmat}                     {\mathbbm{1}}                        
\newcommand{\ii}                        {i}                                  
\renewcommand{\vec}[1]                  {\mathbf{#1}}                        

\providecommand{\eqn}                   {Eq.~}
\providecommand{\eqns}                  {Eqs.~}
\providecommand{\sect}                  {Section~}

\providecommand{\abs}[1]                {\lvert#1\rvert}

\renewcommand{\Im}{\mathop\mathrm{Im}}

\newcommand    {\spinup}     {\mathord{\uparrow}}
\newcommand    {\spindown}   {\mathord{\downarrow}}

\newcommand    {\vk}         {\vec{k}}
\newcommand    {\eF}         {\epsilon_\mathrm{F}} 
\newcommand    {\lambdaISO}  {\lambda_\mathrm{ISO}} 
\newcommand    {\lambdaISOc} {\lambdaISO^\mathrm{c}} 

\newcommand    {\chup}       {\mathcal{C}^{\spinup}}

\newcommand    {\CHup}       {\boldsymbol{\mathcal{C}}^{\spinup}}
\newcommand    {\CHdown}     {\boldsymbol{\mathcal{C}}^{\spindown}}

\newcommand    {\CHspin}     {\boldsymbol{\mathcal{C}}_\text{spin}}
\newcommand    {\sigmaH}     {\sigma_{\mathrm{H}}}
\newcommand    {\sigmaspH}   {\sigma_{\mathrm{SH}}}
\newcommand    {\Tthree}     {\mathrm{T}_3}

\newcommand    {\vecsigma}   {\bm{\sigma}}

\begin{document}
\title{Topological phase transitions driven by next-nearest-neighbor hopping in two-dimensional lattices}
\author{W. Beugeling}
\affiliation{Institute for Theoretical Physics, Utrecht University, Leuvenlaan 4, 3584 CE Utrecht, The Netherlands}
\author{J. C. Everts}
\affiliation{Institute for Theoretical Physics, Utrecht University, Leuvenlaan 4, 3584 CE Utrecht, The Netherlands}
\author{C. \surname{Morais Smith}}
\affiliation{Institute for Theoretical Physics, Utrecht University, Leuvenlaan 4, 3584 CE Utrecht, The Netherlands}

\date{November 19, 2012}

\begin{abstract}
For two-dimensional lattices in a tight-binding description, the intrinsic spin-orbit coupling, acting as a complex next-nearest-neighbor hopping, opens gaps that exhibit the quantum spin Hall effect. In this paper, we study the effect of a real next-nearest-neighbor hopping term on the band structure of several Dirac systems. In our model, the spin is conserved, which allows us to analyze the spin Chern numbers. We show that in the Lieb, kagome, and $\Tthree$ lattices, variation of the amplitude of the real next-nearest-neighbor hopping term drives interesting topological phase transitions. These transitions may be experimentally realized in optical lattices under shaking, when the ratio between the nearest- and next-nearest-neighbor hopping parameters can be tuned to any possible value. Finally, we show that in the honeycomb lattice, next-nearest-neighbor hopping only drives topological phase transitions in the presence of a magnetic field, leading to the conjecture that these transitions can only occur in multigap systems.
\end{abstract}

\maketitle

\section{Introduction}
The study of topological states of matter has received a large increase of interest since the theoretical formulation\cite{KaneMele2005PRL95-22,BernevigEA2006} and experimental observation\cite{KonigEA2007} of the quantum spin Hall effect. Besides the quantum Hall and quantum spin Hall effects in two dimensions, several other topological states of matter have been studied, e.g., topological insulators in three dimensions\cite{HasanKane2010,QiZhang2011} and topological superconductors,\cite{FuKane2008} and they have been classified in a systematic fashion.\cite{Kitaev2009,*RyuEA2010} This classification is based on the notion of topological invariants; in the quantum Hall and spin Hall systems, these invariants are closely related to the quantized Hall and spin Hall conductivities, respectively. These conductivities can be observed when the bulk is in an insulating phase; they are edge properties that exist inside the bulk gaps. Remarkably, these quantities are closely related to the topological invariants of the bulk bands known as the Chern numbers, a property known as the bulk-boundary correspondence. Importantly, these Chern numbers make it possible to classify the topological phases of the system, without having to study the physics on the edges.

Phase transitions between these topological phases are as appealing as the phases themselves. Topological phase transitions are characterized by a change of the topological invariants that classify the system. Because of the topological protection, they can occur only when a gap closes. In that case, the Chern numbers of the energy bands change.\cite{HasanKane2010} Several mechanisms can give rise to these transitions. For instance, in the model by Kane and Mele,\cite{KaneMele2005PRL95-22} it has been shown that by increasing the Rashba spin-orbit coupling one can close the quantum spin Hall gap and open a trivial one. Other types of transitions, between different chiral states, occur in the Hofstadter model, and are driven by variation of the magnetic flux density.\cite{Hatsugai1993PRB} More recently, it has been shown that the intrinsic spin-orbit coupling also drives phase transitions in the honeycomb lattice in the presence of magnetic flux.\cite{GoldmanEA2012,BeugelingEA2012PRB86} In this work, we show that in several two-dimensional lattices topological phase transitions can be driven by tuning the strength of a real next-nearest-neighbor (NNN) hopping term. Although in tight-binding models this term resembles the intrinsic spin-orbit (ISO) coupling (which is a purely imaginary NNN hopping term), it is often neglected, despite the fact that in some systems it may be much stronger than the ISO coupling.

Experiments incorporating ultracold fermionic atoms in optical lattices are ideal candidates for the realization of topological phase transitions, since the strengths of the different hopping parameters can be tuned relatively easily. In the mostly realized square optical lattices, the potential in the $x$ and $y$ directions is usually separable and the NNN hopping (along the diagonal) is zero. For nonseparable optical lattices, the magnitude of the NNN hopping $t'$ is small compared to that of the nearest-neighbor (NN) hopping $t$. However, it has been recently shown that by shaking the optical lattice the effective ratio $t'_\mathrm{eff}/t_\mathrm{eff}$ can be varied in the entire range from $0$ to $\infty$. This occurs because, in a certain regime of parameters, the shaking leads to a renormalization of the hopping, which becomes multiplied by a Bessel function, i.e., $t_\mathrm{eff}=tJ_0(K)$, where $K=V_0/\hbar\omega$ is the shaking parameter, given by the ratio between the amplitude $V_0$ and the frequency $\omega$ of the periodic shaking.\cite{EckardtEA2005,*LignierEA2007,*ZenesiniEA2009,*Hemmerich2010,*StruckEA2012} Moreover, the NNN hopping may get renormalized by a Bessel function with a different argument, thus leading to a full range of possible values for the effective ratio $t'_\mathrm{eff}/t_\mathrm{eff}$.\cite{DiLibertoEA2011}

Here, we focus on tight-binding models, where we assume that all particles loaded onto the lattice are in their respective ground states, and we study topological phase transitions driven by tuning the ratio\footnote{In the experimental setup of a shaken optical lattice, the effective ratio $t'_\mathrm{eff}/t_\mathrm{eff}$ is the relevant tuning parameter.} $t'/t$ in the presence of ISO coupling. In this aspect, we study several lattice geometries. The honeycomb lattice (e.g., of graphene) is one of the most famous systems exhibiting a Dirac dispersion,\footnote{We call systems with a linear dispersion Dirac systems. Strictly speaking, the Hamiltonian of such a system is not necessarily equivalent to the Dirac Hamiltonian, for instance in the case of the Lieb lattice.} where ISO coupling leads to the quantum spin Hall phase. However, the fact that there are only two (spin-degenerate) bands, and hence only one gap, greatly reduces the possible types of topological phase transitions that could occur. In the kagome, $\Tthree$, and Lieb lattices, the spectrum consists of three bands, and hence two gaps, increasing the number of possible topological phase transitions. In this work, we  discuss these three lattices and compare the topological phases and topological phase transitions that they exhibit.

The general tight-binding Hamiltonian, that applies to all lattices presented in this work, reads
\begin{equation}\label{eqn_ham}
  \hat{H}=\hat{H}_{\mathrm{NN}}+\hat{H}_{\mathrm{NNN}}+\hat{H}_{\mathrm{ISO}},
\end{equation}
with
\begin{align}
  \hat{H}_\mathrm{NN} &=-t\sum_{\langle i,j\rangle;\sigma}\hat{s}_{i,\sigma}^\dagger\hat{s}_{j,\sigma},\label{eqn_hnn}\\
  \hat{H}_\mathrm{NNN}&=-t'\sum_{\langle\langle i,j\rangle\rangle;\sigma}\hat{s}_{i,\sigma}^\dagger\hat{s}_{j,\sigma},\label{eqn_hnnn}\\
  \hat{H}_\mathrm{ISO}&=+i\lambdaISO\sum_{\langle\langle i,j\rangle\rangle;\sigma,\sigma'}\hat{s}_{i,\sigma}^\dagger(\vec{e}_{ij}\cdot\vecsigma_{\sigma\sigma'})\hat{s}_{j,\sigma'}.\label{eqn_hiso}
\end{align}
Here, $\hat{s}_{i,\sigma}$ ($\hat{s}^\dagger_{i,\sigma}$) is the annihilation (creation) operator for a particle on site $i$ and with spin $\sigma$. The first two terms $\hat{H}_\mathrm{NN}$ and $\hat{H}_\mathrm{NNN}$ describe NN and NNN hopping, respectively. The third term $\hat{H}_\mathrm{ISO}$ is the ISO coupling; in this term, we have defined the unit vector $\vec{e}_{ij}=(\vec{d}_{ik}\times \vec{d}_{kj}) / |\vec{d}_{ik}\times \vec{d}_{kj}|$ in terms of the bond vectors $\vec{d}_{ik}$ and $\vec{d}_{kj}$, that connects the sites $i$ and $j$ via the unique intermediate site $k$. Finally, $\vecsigma=(\sigma_x,\sigma_y,\sigma_z)$ is the vector of Pauli matrices.

This paper is organized as follows. In \sect\ref{sect_lieb}, we apply the tight-binding model to the Lieb lattice, study the spectrum, and analyze the topological phase transitions using the Chern numbers of the bands. In \sect\ref{sect_kagome}, we repeat this discussion for the kagome and $\Tthree$ lattices. In \sect\ref{sect_honeycomb}, we study a honeycomb lattice  and show that contrarily to the other two examples, no topological phase transitions driven by the real NNN hopping can occur, unless a magnetic field is applied perpendicularly to the lattice. Finally, we discuss the possibilities for experimental realization of this model and we conclude in \sect\ref{sect_discussion}.

\section{The Lieb lattice}
\label{sect_lieb}%
\subsection{Intrinsic spin-orbit coupling and next-nearest-neighbor hopping}%
\begin{figure}[t]
\includegraphics{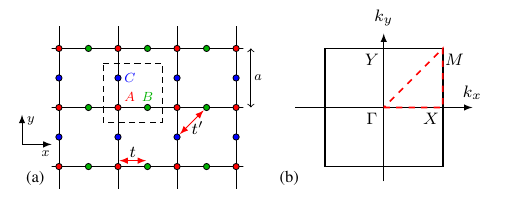}
\caption{(Color online) (a) The geometry of the Lieb lattice. The square unit cell with lattice constant $a\equiv1$ is indicated, together with the three sublattices ($A$, $B$, $C$). An example of a NN (NNN) hopping process $t$ ($t'$) is indicated by the red arrows. (b) First Brillouin zone of the Lieb lattice. The high symmetry points $\Gamma$, $X$, $Y$, and $M$ are given. The red dashed lines indicate the path $\Gamma X M \Gamma$ along which the dispersions are displayed in the subsequent figures.}
\label{fig_lattice}
\end{figure}
First, we investigate various topological phase transitions on the Lieb lattice. In Fig.~\ref{fig_lattice}(a), we show the lattice geometry, with three sites $A$, $B$, and $C$ per unit cell, and lattice constant $a\equiv 1$. The high-symmetry points of the first Brillouin zone (1BZ) are shown in Fig.~\ref{fig_lattice}(b). Both in real space and in reciprocal space, the system satisfies $C_4$ symmetry. By applying a Fourier transformation to the real-space Hamiltonian [\eqns\eqref{eqn_ham}--\eqref{eqn_hiso}], we find the Hamiltonian in momentum space,
\begin{equation}
\hat{H}=\sum_{\vk\in\text{1BZ}}\hat{\Psi}_\vk^\dagger\mathcal{H}_\vk\hat{\Psi}_\vk, \quad \mathcal{H}_\vk=\mathcal{H}_\vk^0\otimes\idmat_{2\times 2}+\mathcal{H}_\vk^\mathrm{ISO}\otimes \sigma_z,
\end{equation}
where $\hat{\Psi}_\vk\equiv(\hat{\Psi}_{\vk,\uparrow},\hat{\Psi}_{\vk,\downarrow})$ with $\hat{\Psi}_{\vk,\sigma}\equiv(\hat{s}_{A,\vk,\sigma},\hat{s}_{B,\vk,\sigma},\hat{s}_{C,\vk,\sigma})$, and
where the $3\times 3$ matrices $\mathcal{H}_\vk^{0}$ and $\mathcal{H}_\vk^\mathrm{ISO}$ are given by
\begin{equation}\label{eqn_hk_nn_nnn}
\mathcal{H}_\vk^0= \begin{pmatrix}
0 & -2tc_x & -2tc_y \\
-2tc_x & 0 & -4t'c_xc_y \\
-2tc_y & -4t'c_xc_y & 0 \end{pmatrix}
\end{equation}
and
\begin{equation}\label{eqn_hk_iso}
\mathcal{H}_\vk^\mathrm{ISO}= 4\ii\lambdaISO \begin{pmatrix}
0 & 0 & 0 \\
0 & 0 & -s_xs_y \\
0 & s_xs_y & 0 \end{pmatrix},
\end{equation}
with $c_\mu=\cos(k_\mu /2)$ and $s_\mu=\sin(k_\mu /2)$ ($\mu=x,y$). The Hamiltonian matrix $\mathcal{H}_\vk$ consists of two uncoupled blocks corresponding to the spin up and spin down projections, related by time-reversal symmetry, i.e., $\mathcal{H}_\vk^\downarrow=(\mathcal{H}_{-\vk}^\uparrow)^*$. Due to the time-reversal and inversion symmetries, the ISO coupling is unable to lift the spin degeneracy. For the calculation of the energy spectrum, it thus suffices to restrict our attention to one spin component, while keeping in mind that the resulting bands are doubly degenerate. In the following, we will restrict ourselves to the spin-up part of the Hamiltonian. We note that this Hamiltonian restricted to a single spin component is formally equivalent to the three-band model presented in Ref.~\onlinecite{HeEA2012} for a copper oxide system.

\begin{figure}[t]
\includegraphics{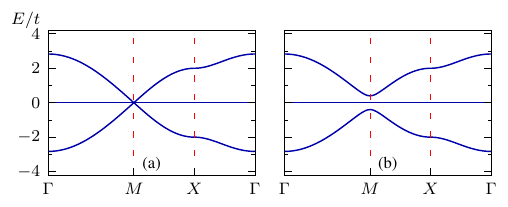}
\caption{\label{fig_isogap}(Color online) (a) The upper and lower bands touch the middle flat band at the $M$ point if $\lambdaISO=t'=0$. (b) A gap appears if the ISO coupling is turned on ($\lambdaISO\neq0$ and $t'=0$).}
\end{figure}

Let us first consider the case where only NN hopping is included ($t'=\lambdaISO=0$). The spectrum then consists of three bands (per spin component), where two dispersing bands touch the flat middle band via a conelike dispersion at the $M$ point; see Fig.~\ref{fig_isogap}(a). There is only one such point in the 1BZ.\cite{GreenEA2010} Since the Nielsen-Ninomiya theorem does not apply here due to the flat band, there is no fermion doubling. \cite{NielsenNinomiya1981,*DagottoEA1986} When the ISO coupling is included ($t'=0$ and $\lambda_\mathrm{ISO}\not=0$), a gap opens at $M$; see Fig.~\ref{fig_isogap}(b). As shown in Refs.~\onlinecite{WeeksFranz2010,GoldmanEA2011}, the resulting gaps are nontrivial, i.e., they exhibit helical edge states, and the upper and lower bands have nonzero Chern numbers. For the definition of the Chern numbers and for an analytic derivation in the case of the Lieb lattice with ISO coupling, we refer to the Appendix. In this case, the result of this computation is $\CHup=(\chup_-,\chup_0,\chup_+)=(1,0,-1)$,
for the Chern numbers of the lower, middle, and upper band, respectively, for the spin-up component. For the spin-down component, we have $\CHdown=-\CHup$. From this property, we immediately obtain the Chern numbers for $\lambdaISO<0$ from those for $\lambdaISO>0$ by interchanging the roles for spin up and spin down. Since the Hamiltonian is diagonal in spin space, we can define the spin Chern number as $\CHspin=\CHup-\CHdown$.\cite{[{In the presence of spin mixing perturbations the collection of Chern numbers for the various bands can be seen as a matrix in spin space, meaning that it is not useful to define a spin Chern number in that case; see }][{}]{ShengEA2006}} Due to the symmetry between the two spin components, it holds that $\CHspin=2\CHup$. By virtue of the bulk-boundary correspondence, the spin Hall conductivity (in units of the spin conductivity quantum $e/4\pi$) equals
\begin{equation}\label{eqn_spinhall}
\sigmaspH =\sum_{\alpha:\ \epsilon_\alpha<\eF}\mathcal{C}_{\text{spin},\alpha},
\end{equation}
where the summation is over the filled bands ($\eF$ is the Fermi energy). The $\mathbb{Z}_2$-index $\nu$ (defined as $0$ for an even and $1$ for an odd number of edge-state pairs) is then related to spin Hall conductivity by $\nu=\sigmaspH/2\mod 2$. An alternative way to derive the Chern numbers and the spin Hall conductivities is by diagonalizing the system in a ribbon geometry, and to count the number of edge states that appear inside the bulk gaps.\cite{Hatsugai1993PRL,BeugelingEA2012PRB86} We remark that for systems that respect inversion symmetry, the $\mathbb{Z}_2$ invariant can also be calculated using the parity eigenvalues at the time-reversal symmetric momenta.\cite{FuKane2007} The advantage of computing the spin Chern numbers, as we do here, is that they are directly related to the quantized spin Hall conductivity in the band gaps.

Now, let us investigate the effects of a purely real NNN hopping ($t'\not=0$ and $\lambda_\mathrm{ISO}=0$). Unlike the ISO coupling, the real NNN term breaks the particle-hole symmetry of the spectrum. This case is, in some sense, more complicated than the previous one, because the Hamiltonian matrix cannot be written in the form $\tilde{\vec{R}}\cdot\tilde{\vec{L}}$, with $\tilde{\vec{L}}$ a three-component vector of matrices that form a basis for a representation of $\mathfrak{su}(2)$. However, it is still possible to extend the components of $\tilde{\vec{L}}$ to a basis of $\mathfrak{su}(3)$. This means that the Chern number does not have a simple interpretation as a winding number any longer. Moreover, the expressions for the eigenvalues are more complicated, although one can still find them analytically.

\begin{figure}[t]
\includegraphics{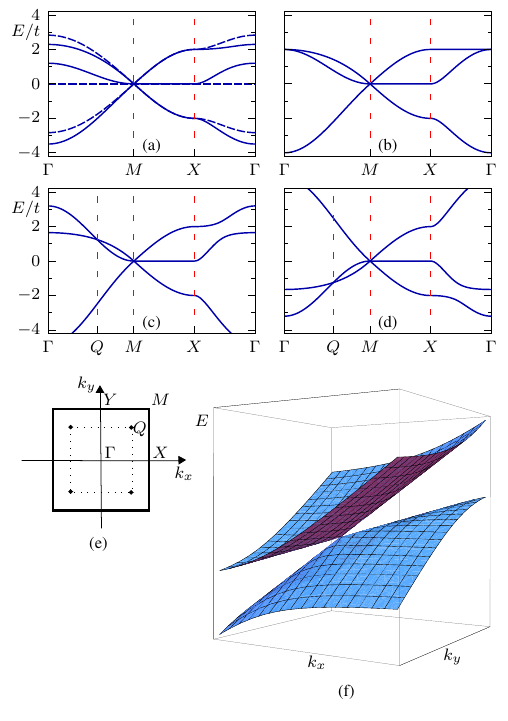}
\caption{\label{fig_realnnn}(Color online) Dispersions for various values of $t'$ along the high symmetry lines in the 1BZ. (a) The case where $t'=0$ is shown by the dashed curves. For $0<t'<0.5t$  the middle band starts to develop a maximum at the $\Gamma$ point (solid curves). (b) For $t'=0.5$ the maximum at the $\Gamma$ point touches the upper band. (c) When $t'>0.5t$ a tilted anisotropic Dirac cone forms at the $Q$ point. (d) The dispersion for $t'<-0.5t$.  In (e), the four corners of the dotted square indicate the positions of the four tilted anisotropic Dirac cones in the 1BZ. (f) A three-dimensional picture of the anisotropic tilted Dirac cone at $Q$.}
\end{figure}

We will now focus on the qualitative behavior of the dispersions as function of $t'$. Let us start with $t'=0$, for which the middle band is flat, as was already discussed before [see Fig.~\ref{fig_realnnn}(a), dashed curves]. For $t'>0$, the middle band starts to develop a maximum at the $\Gamma$ point [see Fig.~\ref{fig_realnnn}(a), solid curves], and ultimately, this maximum touches the upper band for $t'/t=0.5$ [Fig.~\ref{fig_realnnn}(b)]. For $t'/t>0.5$, the middle and upper bands touch at four inequivalent points in the 1BZ of the form $Q=(\pm q, \pm q)$, where $q$ is given by the condition $\cos(qa/2)=t/2t'$ [see Fig.~\ref{fig_realnnn}(c)]. Similarly, the middle and lower bands touch for $t'/t<-0.5$, as shown in Fig.~\ref{fig_realnnn}(d). The dispersion around these touching points $Q$ [see Fig.~\ref{fig_realnnn}(e)] resembles a tilted anisotropic Dirac cone, as shown in Fig.~\ref{fig_realnnn}(f). The positions of the four touching points reflect the $C_4$ symmetry of the lattice. They do not coincide with high symmetry points in the 1BZ. Moreover, we observe that the position of these cones in the 1BZ is tunable by $t'$. In the vicinity of the touching points, the dispersion can be approximated by the generalized Weyl Hamiltonian\cite{GoerbigEA2008}
\begin{equation}\label{eqn_weyl_approx}
  H_\mathrm{eff}(\vec{p})
  = E_0\idmat_{2\times 2} + \tilde{t}\left(\frac{2}{3}p^+\idmat_{2\times 2} + \frac{1}{\sqrt{3}}\sigma_xp^+ + \frac{1}{3}\sigma_yp^-\right),
\end{equation}
where $p^\pm=p_x\pm p_y$ with $\vec{p}=(p_x,p_y)=\vk-Q$ defined as the momentum with respect to the touching point $Q$, $\tilde{t}=t\sqrt{1-(t/2t')^2}$, and $E_0=t/t'$ is the energy of the bands at $Q$. This expression may be found using the Luttinger-Kohn representation\cite{LuttingerKohn1955} of the bands at the band touching points. The tilt comes from the term proportional to $p^+\idmat_{2\times 2}$. We remark that the maximum-tilt condition as defined in Ref.~\onlinecite{GoerbigEA2008} is not satisfied for these cones. In addition, the Hamiltonian \eqref{eqn_weyl_approx} is only of limited interest for our purposes because by itself it does not reproduce the topological properties that we analyze.

The most general case is when both the real and imaginary parts of the NNN hopping are included, $t'\not=0$ and $\lambda_\mathrm{ISO}\not=0$. In this case, the ISO term opens a gap at $M$. Regarding the Chern numbers of the bands, we immediately understand the case $t'<0.5t$: Such a state is adiabatically connected to the case for which $t'=0$, and no gaps close in this regime, so that the Chern numbers of the bands are exactly the same as if one would put $t'=0$. The situation is, however, more subtle for $t'>0.5t$. As mentioned previously, a band touching point occurs at the $\Gamma$ point for $\lambdaISO=0$. Interestingly, this still occurs in the presence of ISO coupling, since its contribution to the Hamiltonian [see \eqn\eqref{eqn_hk_iso}] vanishes at $\Gamma$. Thus, the $\Gamma$ point is protected from the effects of the ISO coupling. Because the upper and middle band do touch when $t'$ is increased to $0.5t$, the Chern numbers of the bands may change. An analytical evaluation of the Chern numbers is in principle possible, but the expressions become tedious to deal with. For this reason, we choose to compute the Chern numbers numerically by resorting to the method proposed by Fukui \emph{et al}.\cite{FukuiEA2005} For the case $t'=0$, the analytical results from a computation similar to the one presented in Ref.~\onlinecite{HeEA2012} coincide with our numerical findings.

\begin{figure}[t]
\includegraphics{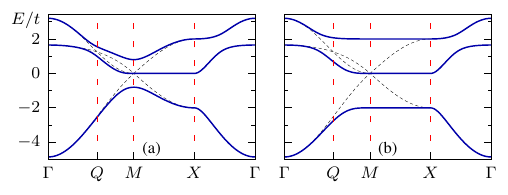}
\caption{(Color online) (a) The black dashed curves indicate the dispersions for $\lambdaISO=0$ and the blue continuous curves indicate the situation for $\lambdaISO\neq 0$. There is no full gap when $\lambdaISO$ is below a critical value $\lambdaISOc$. (b) For a sufficiently large $\lambdaISO>\lambdaISOc$, the spectrum is fully gapped.}
\label{fig_gaptypes}
\end{figure}

Before we present the results from this calculation, we would like to point out a subtlety that arises in the interpretation of the Chern numbers in terms of the spin Hall conductivity. For the latter quantity to be quantized, there must be a full gap in the spectrum, i.e., there must be a range of Fermi energies for which there are only edge states and no bulk states. However, as shown in Fig.~\ref{fig_gaptypes}(a), there may be no full band gap between the middle and upper bands, although they are separated at each point in the 1BZ. In this situation, i.e., if the minimum of the upper band lies at a lower energy value than the maximum of the middle band at a different momentum [$\epsilon_0(\Gamma)\geq\epsilon_+(M)$], we say that the spectrum exhibits a negative indirect gap. The Chern numbers of the bands are well defined, since they do not touch anywhere, and no topological phase transition occurs.
In this example, a $2/3$-filled system will always have both the upper band and the middle band partially filled, which would classify the bulk as a semimetal, preventing the helical edge states to be observed. In the semimetallic regime, i.e., with partially filled bands, the spin Hall conductivity is not quantized.\cite{SunFradkin2008} For $1/3$ filling, this problem does not occur since in this case there is a full band gap. At this filling, the system behaves as an insulator and the spin Hall conductivity carried by the helical edge states can be experimentally observed. By virtue of symmetry, a similar reasoning occurs for $t'<0$, but with the roles of the $1/3$ and $2/3$ filling, and of the lower and upper bands interchanged. Finally, the spectrum is fully gapped if the ISO coupling is sufficiently strong, i.e., there is a critical value $\lambdaISOc(t')$, such that the system exhibits only full gaps if the ISO coupling satisfies $\abs{\lambdaISO}>\lambdaISOc$; see Fig.~\ref{fig_gaptypes}(b).

\begin{figure}[t]
\centering
\includegraphics[width=86mm]{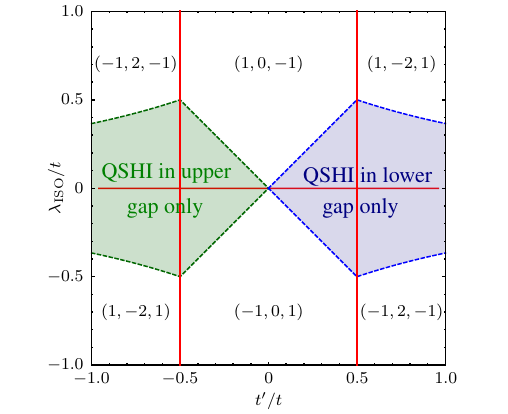}
\caption{\label{fig_phasediagram}(Color online) Phase diagram as function of $\lambdaISO$ and $t'$. The red (solid) lines indicate positions for which a band gap closes, meaning that the Chern numbers can change. In the regions enclosed by these lines the various (spin) Chern numbers $\CHup=\CHspin/2$ are indicated. The regions delimited by the blue (light gray) or green (dark gray) dashed lines indicate the parameter regimes for which there is no full gap between the middle and upper bands or middle and lower bands, respectively.}
\end{figure}

In Fig.~\ref{fig_phasediagram}, we display the phase diagram  in the $(t',\lambdaISO)$ plane, where we indicate the regions with distinct Chern numbers for the bands. The topological phase transitions, i.e., the values $(t',\lambdaISO)$ where band touchings occur, are indicated by solid lines. Moreover, we have shaded the parameter regimes in blue (green) where the middle and upper (lower) bands overlap in energy and the only full gap is at $1/3$ ($2/3$) filling. In all open gaps, the system behaves as a quantum spin Hall insulator with spin Hall conductivity equal to $\sigmaspH=\pm1$. For $\abs{t'/t}<0.5$, the conductivities in the lower ($1/3$ filling) and upper ($2/3$ filling) gap are equal, since the middle band has zero Chern number. For $\abs{t'/t}>0.5$, the two bands have opposite conductivities.

Every phase transition in the phase diagram may be understood in terms of the difference in the Chern numbers of the phases at both sides of the transition. For the transitions defined by the lines $t'/t=\pm0.5$, either the lower or the upper gap closes, and the system behaves as a metal at the gap closing energy; the other gap is still helical. For example, if $t'/t=0.5$ and $\lambdaISO>0$, the difference of the Chern numbers is $\Delta\CHup=(1,0,-1)-(1,-2,1)=(0,2,-2)$. We observe that the change of the Chern numbers of the two touching bands is $\pm 2$. This value can be understood from the fact that the bands touch at $\Gamma$, and they behave quadratically around this point, i.e., there is a Berry phase of $4\pi$ associated to this touching point. For the transitions at $\lambdaISO=0$ (i.e., between the topological phases for $\lambdaISO>0$ and $\lambdaISO<0$), we distinguish two cases. First, for $\abs{t'/t}<0.5$, the three bands touch at $M$, and the Chern-number difference between the two phases is equal to $\Delta\CHup=(1,0,-1)-(-1,0,1)=(2,0,-2)$, i.e., $\pm2$ for the upper and lower bands, respectively. Secondly, for $t'/t>0.5$, the three bands touch at $M$ as before, but also at the four Weyl cones. The change in Chern numbers is therefore $\Delta\CHup=(1,-2,1)-(-1,2,-1)=(2,-4,2)$, with the contributions $(2,0,-2)$ from the $M$ point and $(0,-4,4)$ from the four Weyl points, respectively. The transition at $\lambdaISO=0$ for $t'/t<-0.5$ is similar.

\subsection{The Lieb lattice with a dimerization term}
In Ref.~\onlinecite{WeeksFranz2010}, it has been shown that a dimerization term, defined by changing the hopping amplitude $t$ in \eqn\eqref{eqn_hnn} alternatingly to $t\pm\alpha$, leads to a trivial gap. In this section, we investigate the effects of the dimerization term in combination with the other terms in the Hamiltonian~\eqref{eqn_ham}. The Hamiltonian then reads
\begin{equation}\label{eqn_hk_with_dimer}
\mathcal{H}_\vk=\mathcal{H}_\vk^0\otimes\idmat_{2\times 2}+\mathcal{H}_\vk^\mathrm{ISO}\otimes \sigma_z+ \mathcal{H}_\vk^\mathrm{dim}\otimes\idmat_{2\times2},
\end{equation}
where the dimerization term is given by
\begin{equation}\label{eqn_hk_dimer}
\mathcal{H}_\vk^\mathrm{dim}= \begin{pmatrix}
0 & -2\ii\alpha s_x & -2\ii\alpha s_y \\
2\ii\alpha s_x & 0 & 0 \\
2\ii\alpha s_y & 0 & 0 \end{pmatrix}
\end{equation}
in momentum space. This term breaks the $C_4$ symmetry of the lattice. In principle, the signs before $s_y$ may be chosen oppositely, but this choice does not affect the following results qualitatively, since it amounts merely to the transformation $(k_x,k_y)\to(k_x,-k_y)$ in momentum space.

\begin{figure}[b]
\includegraphics[width=86mm]{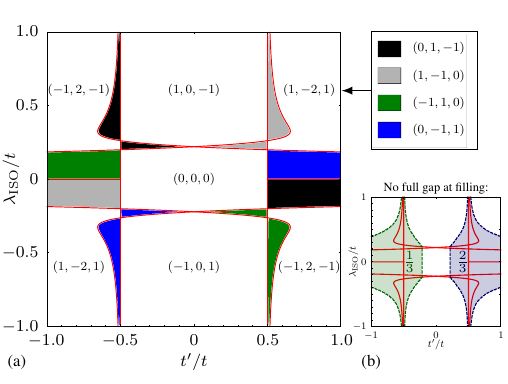}
\caption{\label{fig_phasediagram_dimer}(Color online) (a) Phase diagram as function of $\lambdaISO$ and $t'$ for a nonzero dimerization term ($\alpha/t=0.3$). We indicate the phase transitions by red (solid) lines, and the Chern numbers of the bands $\CHup=\CHspin/2$ are indicated in the regions delimited by these lines. In the shaded regions, one gap is helical and the other is trivial. (b) The regimes with indirect negative gaps (semimetallic behavior) at $1/3$ or $2/3$ filling are indicated by the green (dark gray) and blue (light gray) regions. In the white regions, the system has full band gaps at $1/3$ and $2/3$ filling.}
\end{figure}

The effect of the dimerization  for $\alpha/t=0.3$ on the phase diagram is shown in Fig.~\ref{fig_phasediagram_dimer}(a). We observe that, compared to the case for $\alpha=0$ in Fig.~\ref{fig_phasediagram}, there are additional phase transitions; in particular, the existing phase transitions have been split up. Comparing Figs.~\ref{fig_phasediagram} and \ref{fig_phasediagram_dimer}(a), we observe that in both cases there are phase transitions at $t'/t=\pm0.5$.
However, for $\alpha\not=0$, there is an additional phase transition close to the one at $t'/t=\pm0.5$ (for $\abs{\lambdaISO/t}\gtrsim0.2$). Let us compare the difference in Chern numbers at these transitions as before: For $\alpha=0$, this difference is $\Delta\CHup=(1,0,-1)-(1,-2,1)=(0,2,-2)$. With nonzero $\alpha$, there are two transitions, from $(1,0,-1)$ to $(1,-1,0)$ and  from $(1,-1,0)$ to $(1,-2,1)$; thus, the dimerization term has created an intermediate phase with $\CHup=(1,-1,0)$, which has a helical structure in the lower gap (unaffected by the transition), and a trivial upper gap. The first transition [from $(1,0,-1)$ to $(1,-1,0)$ at $\abs{t'/t}=0.5$] occurs due to a band touching at the point $\Gamma$; expanding the bands around this point, we observe that they are quadratic along one diagonal [$\vk=(q,q)$] and linear along the other [$\vk=(q,-q)$]. The other phase transition [from $(1,-1,0)$ to $(1,-2,1)$ for $t'/t$ slightly larger than $0.5$] occurs due to a band touching involving a single Weyl cone on the diagonal $\vk=(q,q)$. Thus, the original transition with a Chern-number difference of $\pm2$ for the touching bands has been split up in two transitions with a Chern-number difference of $\pm1$.

The transition at zero ISO coupling, originally between the phase with $(1,0,-1)$ and $(-1,0,1)$ has also been split up by the dimerization term. Going from $\lambdaISO/t=0.3$ to $\lambdaISO/t=-0.3$ for $0<\abs{t'/t}<0.5$ and $\alpha/t=0.3$, one encounters four gap closures (and two for $t'=0$), where the bands (the middle and either the upper or lower) touch in one point (in a Weyl cone-like behavior). At $\lambdaISO=0$ there is no transition, i.e., the system remains trivially gapped. For $t'/t>0.5$, the dimerization splits the transition originally at $\lambdaISO=0$ into three: For $\alpha>0$, the middle and upper bands touch at two Weyl cone-like points if $\lambdaISO=0$, and the corresponding change of Chern numbers is $\pm2$. At $\lambdaISO/t\approx\pm0.2$ (for $\alpha/t=0.3$), the middle and lower bands touch at one point, accounting for a change of Chern numbers of $\pm1$.

The dimerization term also modifies the parameter regimes where one of the gaps is a negative indirect gap, as may be observed by comparing the shaded regions of Figs.~\ref{fig_phasediagram} and \ref{fig_phasediagram_dimer}(b) (without and with dimerization term, respectively). There are two qualitative differences: First, with the dimerization term, the critical spin-orbit coupling $\lambdaISOc(t')$, tends to infinity for $\abs{t'/t}\to0.5$, whereas $\lambdaISOc(t')$ remains finite if $\alpha=0$. Secondly, with dimerization, there are no negative indirect gaps if the NNN hopping amplitude satisfies $\abs{t'}<\abs{\alpha}/\sqrt{2}$ (i.e., $\abs{t'/t}<0.21$, approximately, for $\alpha/t=0.3$). We remark that for all topological phases indicated in Fig.~\ref{fig_phasediagram_dimer}(a), there is a regime of parameters $(t',\lambdaISO)$ for which it may be observed in a system with two full band gaps, i.e., such that the spin Hall conductivity is quantized in both gaps.

\section{The kagome and $\Tthree$ lattices}
\label{sect_kagome}%
\begin{figure}[t]
\includegraphics{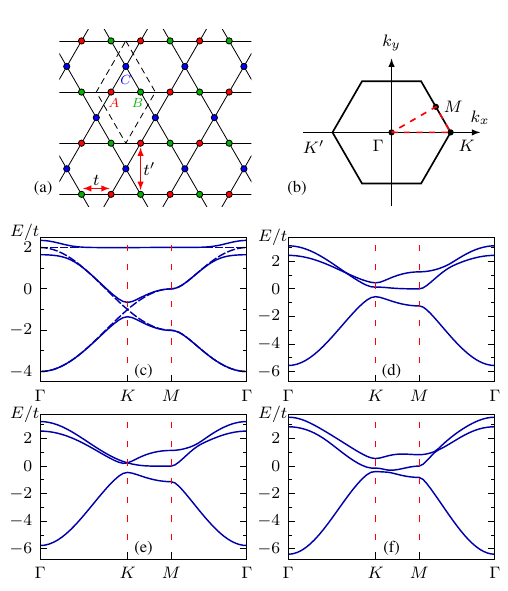}  
\caption{\label{fig_kagome}(Color online) (a) The kagome lattice with the three inequivalent sites $A$, $B$, and $C$ inside the unit cell (dashed area). The NN and NNN hoppings are indicated by the red arrows. (b) First Brillouin zone of the kagome lattice with high-symmetry points. (c) Dispersions along contour $\Gamma K M \Gamma$ for $\lambdaISO=0$ and $t'=0$ (dashed curves) and for $\lambdaISO/t=0.1$ and $t'=0$ (solid curves). (d)--(f) Dispersions for $\lambdaISO/t=0.1$ and $t'/t=0.39$, $t'/t=0.443$, and $t'/t=0.6$, respectively. The two upper bands touch at $K$ for $t'/t\approx0.443$.}
\end{figure}%
The existence of topological phase transitions driven by the NNN hopping $t'$ in the Lieb lattice raises the question whether they are specific to that lattice or whether they would also appear in other lattices. In this section, we first investigate the kagome lattice, which has the lattice structure shown in Fig.~\ref{fig_kagome}(a). The unit cell consists of three sites, but these all have the same number of NNs, unlike the Lieb lattice. From a Fourier transformation of the real-space Hamiltonian \eqref{eqn_ham}--\eqref{eqn_hiso} for this particular lattice geometry, we find the Hamiltonian in momentum space [for the structure of the 1BZ, see Fig.~\ref{fig_kagome}(b)], which is given by
\begin{equation}
\mathcal{H}_\vk^\mathrm{K}
 = \mathcal{H}^\mathrm{K,NN}_\vk\otimes\idmat_{2\times 2}
  +\mathcal{H}^\mathrm{K,NNN}_\vk\otimes\idmat_{2\times 2}
  +\mathcal{H}_\vk^\mathrm{K,ISO}\otimes \sigma_z,
\end{equation}
in the basis $\hat{\Psi}_\vk=(\hat{\Psi}_{\vk,\uparrow},\hat{\Psi}_{\vk,\downarrow})$, where $\hat{\Psi}_{\vk,\sigma}=(\hat{s}_{A,\vk,\sigma},\hat{s}_{B,\vk,\sigma},\hat{s}_{C,\vk,\sigma})$.
The $3\times 3$ matrices $\mathcal{H}^\mathrm{K,NN}_\vk$, $\mathcal{H}^\mathrm{K,NNN}_\vk$, and $\mathcal{H}_\vk^\mathrm{K,ISO}$ are given by
\begin{align}
\mathcal{H}^\mathrm{K,NN}_\vk&= -2t\begin{pmatrix}
0 & c_1 & c_3 \\
c_1 & 0 & c_2 \\
c_3 & c_2 & 0 \end{pmatrix},\label{eqn_hk_kagome_nn}\\
\mathcal{H}^\mathrm{K,NNN}_\vk&= -2t'\begin{pmatrix}
0 & b_1 & b_3 \\
b_1 & 0 & b_2 \\
b_3 & b_2 & 0 \end{pmatrix},\label{eqn_hk_kagome_nnn}\\
\intertext{and}
\mathcal{H}_\vk^\mathrm{K,ISO}&= -2\ii\lambdaISO\begin{pmatrix}
0 & b_1 & -b_3 \\
-b_1 & 0 & b_2 \\
b_3 & -b_2 & 0 \end{pmatrix},\label{eqn_hk_kagome_iso}
\end{align}
where $c_i=\cos(\vk\cdot\vec{e}_i)$ and $b_i=\cos(\vk\cdot\vec{f}_i)$ are given in terms of the NN vectors $\vec{e}_1=(1,0)$, $\vec{e}_2=(-1,\sqrt{3})/2$ and $\vec{e}_3=(-1,-\sqrt{3})/2$, and the NNN vectors $\vec{f}_1=\vec{e}_2-\vec{e}_3$, $\vec{f}_2=\vec{e}_3-\vec{e}_1$, and $\vec{f}_3=\vec{e}_1-\vec{e}_2$.

\begin{figure}[t]
\includegraphics{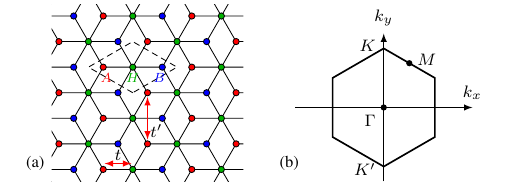}
\caption{\label{fig_t3}(Color online) (a) The $\Tthree$ lattice with the three inequivalent sites $A$, $H$, and $B$ inside the unit cell (dashed area). The NN and NNN hoppings are indicated by the red arrows. (b) First Brillouin zone of the $\Tthree$ lattice with high-symmetry points.}
\end{figure}

In absence of NNN hopping and ISO coupling, the spectrum consists of three bands: Two dispersing bands, which resemble the dispersion of graphene, with two inequivalent Dirac points inside the 1BZ, where the bands touch at $E/t=-1$, and a flat band at $E/t=2$, which touches the middle band at $\Gamma$;\cite{GuoFranz2009} see Fig.~\ref{fig_kagome}(c), dashed lines. If only real NNN hopping is included (i.e., $t'\not=0$ and $\lambdaISO=0$), then no gaps open between the bands. If only ISO coupling is included [see the solid lines in Fig.~\ref{fig_kagome}(c)], then gaps exhibiting helical states appear. Unlike the Lieb lattice, varying the ISO coupling drives several phase transitions, not just one for $\lambdaISO=0$. For the case that both the real and imaginary parts of the NNN hopping are nonzero ($t'\not=0$ and $\lambdaISO\not=0$), we find many band touching points, see Figs.~\ref{fig_kagome}(d)--\ref{fig_kagome}(f). For instance, one of the gaps closes at $K$ and $K'$ for $\lambdaISO/t=\pm\sqrt{3}(t'/t-1/2)$. Additional gap transitions occur due to bands touching at points between $\Gamma$ and $K$/$K'$, between $\Gamma$ and $M$, and between $M$ and $K$/$K'$. Without explicit computation of the Chern numbers, we can use the symmetry properties of the 1BZ to predict the change of Chern number in these transitions. The band touching points at $\Gamma$, at $K$/$K'$, between $\Gamma$ and $K$/$K'$, between $\Gamma$ and $M$, and between $M$ and $K$/$K'$ have multiplicities $1$, $2$, $6$, $6$, and $6$, respectively. The difference in the Chern numbers of the bands between the phases at both sides of the corresponding transitions is an integer multiple of this multiplicity. Although a phase diagram like Fig.~\ref{fig_phasediagram} can be made also for the kagome lattice, it is outside the scope of this paper. The question of whether topological phase transitions driven by real NNN  appear in other two-dimensional Dirac-like systems than the Lieb lattice has been answered positively.

In order to check how generic this behavior is, we attempt to reproduce it in more lattice geometries. The lattice structure shown in Fig.~\ref{fig_t3}(a) is known as the $\Tthree$ lattice.\cite{BerciouxEA2009,BerciouxEA2011} In momentum space, the Hamiltonian for this lattice can be written as
\begin{equation}
\mathcal{H}_\vk^{\Tthree}
 = \mathcal{H}^{\Tthree\mathrm{,NN}}_\vk\otimes\idmat_{2\times 2}
  +\mathcal{H}^{\Tthree\mathrm{,NNN}}_\vk\otimes\idmat_{2\times 2}
  +\mathcal{H}_\vk^{\Tthree\mathrm{,ISO}}\otimes \sigma_z,
\end{equation}
in the basis $\hat{\Psi}_\vk=(\hat{\Psi}_{\vk,\uparrow},\hat{\Psi}_{\vk,\downarrow})$, where $\hat{\Psi}_{\vk,\sigma}=(\hat{s}_{A,\vk,\sigma},\hat{s}_{H,\vk,\sigma},\hat{s}_{B,\vk,\sigma})$, with
\begin{align}
\mathcal{H}^{\Tthree\mathrm{,NN}}_\vk&= -t\begin{pmatrix}
0 & g^*(\vk) & 0\\
g(\vk) & 0 & g^*(\vk) \\
0 & g(\vk) & 0 \end{pmatrix},\label{eqn_hk_t3_nn}\\
\mathcal{H}^{\Tthree\mathrm{,NNN}}_\vk&= -t'\begin{pmatrix}
h(\vk) +h^*(\vk) & 0 & 0 \\
0 & 0 & 0 \\
0 & 0 & h(\vk) +h^*(\vk) \end{pmatrix},\label{eqn_hk_t3_nnn}\\
\intertext{and}
\mathcal{H}_\vk^{\Tthree\mathrm{,ISO}}&= -\ii\lambdaISO\begin{pmatrix}
h(\vk) -h^*(\vk) & 0 & 0 \\
0 & 0 & 0 \\
0 & 0 & -h(\vk) +h^*(\vk) \end{pmatrix},\label{eqn_hk_t3_iso}
\end{align}
where we define $g(\vk)=\ee^{\ii\vk\cdot\vec{e}_1}+\ee^{\ii\vk\cdot\vec{e}_2}+\ee^{\ii\vk\cdot\vec{e}_3}$ and $h(\vk)=\ee^{\ii\vk\cdot\vec{f}_1}+\ee^{\ii\vk\cdot\vec{f}_2}+\ee^{\ii\vk\cdot\vec{f}_3}$, with $\vec{e}_i$ and $\vec{f}_i$ the same as given for the kagome lattice. We remark that there are other possibilities for real NNN terms as well, but we restrict ourselves to the one given here, because it models hopping between the same pairs of sites as the ISO coupling does, namely between two $A$ sites or two $B$ sites. It has been proved earlier that the ISO coupling leads to the opening of gaps between the middle (flat) band and the two other bands, both exhibiting two helical edge-state pairs.\cite{BerciouxEA2011} With inclusion of the real NNN term given by \eqn\eqref{eqn_hk_t3_nnn}, it may be derived analytically that the gap between the middle band and one of the other bands closes at the BZ points $K$ and $K'$ [see Fig.~\ref{fig_t3}(b)] when $\abs{t'/t}=\sqrt{3}\abs{\lambdaISO/t}$. At such a transition, the Chern numbers of the three bands typically change from $(2,0,-2)$ to $(0,2,-2)$. Thus, the $\Tthree$ lattice also accommodates topological phase transitions driven by real NNN hopping. Next, we also investigate the honeycomb lattice, aiming at checking the prerequisites for this behavior.

\section{The honeycomb lattice}
\label{sect_honeycomb}
\begin{figure}[t]
\includegraphics{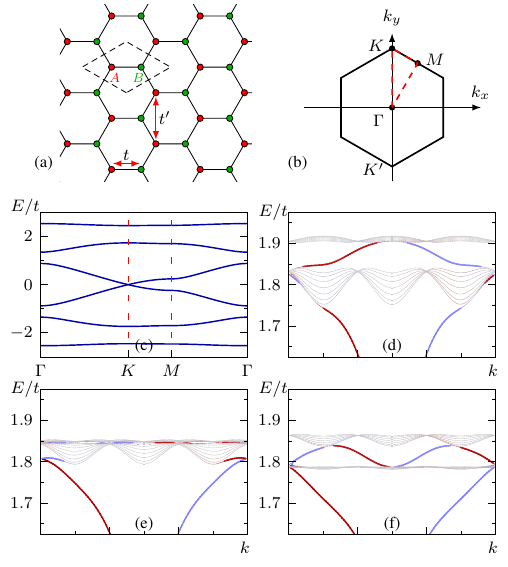}
\caption{\label{fig_honeycomb}(Color online) (a) The honeycomb lattice with the two inequivalent sites $A$ and $B$ inside the unit cell (dashed area). The NN and NNN hoppings are indicated by the red arrows. (b) First Brillouin zone of the honeycomb lattice with high-symmetry points. (c) Dispersions for $\lambdaISO=0$, $t'=0$, and $1/3$ flux quantum per unit cell, along the contour $\Gamma K M \Gamma$ of the magnetic Brillouin zone ($1/3$ of the size of the 1BZ). (d)--(f) Edge-state plots in the energy regime of the two highest bands for $\lambdaISO/t=0$ and $t'/t=-0.18$, $t'/t=-0.2$, and $t'/t=-0.22$, respectively. For these plots, the dispersions have been computed for the system in a cylindrical geometry (see Ref.~\onlinecite{BeugelingEA2012PRB86}). The light gray curves are the bulk bands. The red (dark gray) and blue (medium gray) curves are the edge states on the two opposite edges of the cylinder. The variation of $t'$ closes the gap between these bands, and changes the Hall conductivity in this gap from $-2e^2/h$ to $4e^2/h$.}
\end{figure}

The honeycomb lattice [see Figs.~\ref{fig_honeycomb}(a) and \ref{fig_honeycomb}(b)] is one of the most intensively studied Dirac systems because it models the electronic structure of graphene very well. Since the work by Kane and Mele,\cite{KaneMele2005PRL95-14,KaneMele2005PRL95-22} it is known that ISO coupling in the honeycomb lattice gives rise to the quantum spin Hall phase. The real NNN hopping has been neglected in this model, although it is several orders of magnitude stronger than the ISO coupling: In pure graphene, $\abs{t'/t}\sim 0.1$ and $\lambdaISO/t\sim10^{-6}$--$10^{-4}$.\cite{BoettgerTrickey2007,*VanGelderenMoraisSmith2010} Here, our aim is to study the dispersions and the possibilities for phase transitions if both ISO coupling and real NNN hopping are included in the tight-binding model for the honeycomb lattice.

The Hamiltonian in momentum space for the honeycomb lattice is given by
\begin{equation}
\mathcal{H}_\vk^\mathrm{H}
 = \mathcal{H}^\mathrm{H,NN}_\vk\otimes\idmat_{2\times 2}
  +\mathcal{H}^\mathrm{H,NNN}_\vk\otimes\idmat_{2\times 2}
  +\mathcal{H}_\vk^\mathrm{H,ISO}\otimes \sigma_z,
\end{equation}
in the basis $\hat{\Psi}_\vk=(\hat{\Psi}_{\vk,\uparrow},\hat{\Psi}_{\vk,\downarrow})$, where $\hat{\Psi}_{\vk,\sigma}=(\hat{s}_{A,\vk,\sigma},\hat{s}_{B,\vk,\sigma})$.
The $2\times 2$ matrices $\mathcal{H}_\vk^\mathrm{H,NN}$, $\mathcal{H}_\vk^\mathrm{H,NNN}$, and $\mathcal{H}_\vk^\mathrm{H,ISO}$ are given by
\begin{align}
\mathcal{H}_\vk^\mathrm{H,NN}&= -t\begin{pmatrix}
0 & g(\vk) \\
g^*(\vk) & 0
\end{pmatrix},\label{eqn_hk_honeycomb_nn}\\
\mathcal{H}_\vk^\mathrm{H,NNN}&= -t'\begin{pmatrix}
h(\vk) +h^*(\vk) & 0\\
  0  & h(\vk) +h^*(\vk)
\end{pmatrix},\label{eqn_hk_honeycomb_nnn}\\
\intertext{and}
\mathcal{H}_\vk^\mathrm{H,ISO}&= -\ii\lambdaISO\begin{pmatrix}
h(\vk) -h^*(\vk) & 0\\
  0 &  -h(\vk) +h^*(\vk)
\end{pmatrix},\label{eqn_hk_honeycomb_iso}
\end{align}
where $g(\vk)=\ee^{\ii\vk\cdot\vec{e}_1}+\ee^{\ii\vk\cdot\vec{e}_2}+\ee^{\ii\vk\cdot\vec{e}_3}$ and $h(\vk)=\ee^{\ii\vk\cdot\vec{f}_1}+\ee^{\ii\vk\cdot\vec{f}_2}+\ee^{\ii\vk\cdot\vec{f}_3}$ as defined in the previous section. For this model, it is straightforward to derive the dispersion of the two bands analytically, as
\begin{equation}\label{eqn_honeycomb_dispersion}
  E=t'[h(\vk)+h^*(\vk)]\pm\sqrt{\abs{t g(\vk)}^2+\lambdaISO^2[\ii h(\vk)-\ii h^*(\vk)]^2}.
\end{equation}
The bands touch if and only if the square root equals zero. Since both terms in this square root are positive, the bands touch only when $g(\vk)=0$ (i.e., at $\vk=K$ or $\vk=K'$) and $\lambdaISO \Im h(\vk)=0$; because $\Im h(K)\not=0\not=\Im h(K')$, the bands can only touch for $\lambdaISO=0$. The real NNN hopping has no effect on the touching of the bands; it merely shifts the energies of both bands up or down. Thus, the conclusion is that, in the honeycomb lattice, there are no phase transitions driven by the real NNN coupling. The only topological phase transition in this model is the one at $\lambdaISO=0$, which forms the boundary between two helical phases with opposite signs of the spin Hall conductivity.

However, for the honeycomb lattice in a magnetic field, the results are different. (For the framework of a honeycomb lattice in a magnetic field, see, e.g.,\ Refs.~\onlinecite{GoldmanEA2012,BeugelingEA2012PRB86}.) The Chern numbers $\CHup$ and $\CHdown$ are well defined because of the absence of a coupling between the spin components, but they do not satisfy the condition $\CHup=-\CHdown$, which applies only if there is time-reversal symmetry.

In the case that ISO coupling is absent, there are no terms that break the spin symmetry, so that one observes spin-degenerate \emph{chiral} (i.e., quantum Hall) phases only. Due to the spin degeneracy, the Chern numbers satisfy $\CHup=\CHdown$, and the Hall conductivity is quantized in steps of $2e^2/h$. In Fig.~\ref{fig_honeycomb}(c), we have displayed the band structure for the honeycomb lattice under a flux of $1/3$ flux quantum per unit cell. In this case, there are six bands per spin component, rather than two. Here, in absence of ISO coupling, the variation of the real NNN hopping induces phase transitions between chiral states. In Figs.~\ref{fig_honeycomb}(d)--\ref{fig_honeycomb}(f), we show that the gap between the two highest bands closes when varying $t'/t$. In Figs.~\ref{fig_honeycomb}(d) and (f), there are one and two edge-state pairs in the relevant gap, respectively. In addition, the direction of the edge states is flipped. Thus, the Hall conductivity inside this gap changes from $\sigmaH=-2$ to $\sigmaH=4$ (in units of $e^2/h$; recall that the spin degeneracy yields an additional factor of $2$). Here, we note that the change of $6$ is explained from the fact that there are three inequivalent band touching points in the 1BZ due to the fact that there is $1/3$ flux quantum per unit cell, and from the twofold spin degeneracy.

In presence of ISO coupling together with the magnetic field, more exotic phase transitions can be expected. In that case, the spin degeneracy is lifted and the condition $\CHup=\pm\CHdown$ is generally not satisfied. As a consequence, more types of topological phases can be observed than the helical and spin-degenerate chiral phases, e.g., chiral phases which are not spin degenerate, and even phases which are neither chiral nor helical, but a mixture of both. In Refs.~\onlinecite{GoldmanEA2012,BeugelingEA2012PRB86}, it has been shown that variation of the ISO coupling leads to a topological phase transition between a chiral and a helical state, as well as several transitions between other types of topological phases. If additionally real NNN hopping is present, this picture remains qualitatively unaltered. In conclusion, topological phase transitions in the honeycomb lattice subjected to a perpendicular magnetic field can be driven by ISO coupling as well as by real NNN hopping. This conclusion is not only valid for the honeycomb lattice in a magnetic field, but also for the Lieb, kagome, and $\Tthree$ (among other) lattices. In case of a magnetic field, there are chiral phases.\cite{BerciouxEA2011,LiuEA2012} If the magnetic field is combined with the ISO coupling, also helical phases and nonchiral-nonhelical phases appear. As in the case of the honeycomb lattice, there are topological phase transitions between them, driven either by ISO coupling or by real NNN hopping.

\section{Discussion}
\label{sect_discussion}

The theoretical results presented above raise the question whether the transitions driven by real NNN hopping appear in an experimentally accessible regime. We envisage systems with cold atoms in an optical lattice to be the most promising candidates for observation of these phase transitions, because of the possibility of tuning the hopping parameters. In a static optical lattice, $t'/t$ is typically of the order of $10^{-2}$,\cite{DiLibertoEA2011} far from the transitions that happen at $t'/t=0.5$. However, shaken optical lattices exhibit renormalized hopping coefficients, where the bare values become multiplied by a Bessel function.\cite{EckardtEA2005,*LignierEA2007,*ZenesiniEA2009,*Hemmerich2010,*StruckEA2012} In the most studied case, one applies a periodic shaking to the lattice with the effective potential $V=V_0\cos(\omega\tau)\sum_{i,j}(i+j)\hat{n}_{ij}$, where $V_0$ is the amplitude and $\omega$ is the frequency of the shaking. Here, $(i,j)$ labels the lattice sites in the two-dimensional lattice, $\tau$ is real time, and $\hat{n}_{ij}$ denotes the density operator at site $(i,j)$. In the high-frequency limit $\hbar\omega\gg V_0,t$, the effective hopping may be determined by applying Floquet theory. One finds that $t_\mathrm{eff}=tJ_0(K)$, where the shaking parameter $K=V_0/\hbar\omega$. Moreover, the argument of the Bessel function generally differs for NN and NNN hopping (see, e.g., Ref.~\onlinecite{DiLibertoEA2011} for the case of a non-separable square lattice), thus allowing one to tune the ratio $t'_\mathrm{eff}/t_\mathrm{eff}$ in the entire range of possibilities ($0$ to $\infty$). Recently, it has been demonstrated that in the shaken kagome lattice, the band parameters can be controlled such that band touching points are reached.\cite{HaukeEA2012}

The detection of topological phases in optical-lattice experiments presents a challenge because it is impossible to perform (Hall) conductivity measurements in such systems. First of all, the atoms are neutral, so that there are no electric currents. Therefore, different detection methods must be envisaged to image the edge states and to determine the topological properties. Recently, there has been a proposal to detect chiral edge states in a Hofstadter optical lattice by using angular-momentum sensitive Bragg spectroscopy.\cite{GoldmanEA2012PRL} Additionally, the edge states are probed in a clever way using Raman transitions, in order to isolate the edge-state signal from the background. A second potential problem is the harmonic trap in which the atoms are confined. The lack of sharp edges means that the notion of edge states is no longer well defined. This phenomenon is visible in the spectrum of a system in a harmonic background potential: The ordered structure of the spectrum, consisting of the bulk bands and a few edge states in the gaps, is destroyed by this potential.\cite{BuchholdEA2012} However, it has been shown that the topological invariants remain well defined in the case of a system with soft edges. In other words, sharp edges are not necessary in order to observe the topological phases.\cite{BuchholdEA2012,GoldmanEA2012PRL}

Besides the more complex optical lattices with multisite unit cells considered here, interesting topological phases may also occur in simpler square lattices loaded with bosons in higher excited states, e.g., $p$, $d$, and $f$ orbitals (see Refs.~\onlinecite{WirthEA2011}, \onlinecite{OlschlagerEA2012}, and \onlinecite{OlschlagerEA2011}, respectively). Indeed, for a square superlattice with deeper and shallower wells arranged on a checkerboard pattern, it is possible to realize relatively long-lived Bose-Einstein condensates in the deeper wells which are locally in a metastable $p_x$ or $p_y$ state, whereas the ones in the shallower wells are in an $s$-orbital state. The single-particle spectrum of this system is then very similar to the one of the Lieb lattice, which is understood from the formal equivalence of the Hamiltonians involving three orbitals per unit cell.\cite{WirthEA2011}

In conclusion, we have shown that in several Dirac-type tight-binding models, it is possible to drive topological phase transitions with the real NNN hopping. In the Lieb, kagome, and $\Tthree$ lattices, it drives transitions between helical phases with different values of the spin Hall conductivity. This is not true for the honeycomb lattice in zero magnetic field. In a magnetic field, the phase transitions are possible. In this aspect, we conjecture that it is necessary to have at least two gaps (three bands) in order for these phase transitions to appear. Our results could be observed with state-of-the-art shaken optical lattices loaded with fermions.

\acknowledgments
We thank M. O. Goerbig and V. Juri\v{c}i\'{c} for useful discussions. This work was supported by the Netherlands Organisation for Scientific Research (NWO).

\appendix*

\section{Evaluation of the Chern numbers}
\label{sect_app_chern}
By virtue of the bulk-boundary correspondence,\cite{Hatsugai1993PRB,Hatsugai1993PRL} the Chern numbers of the bands yield information about the (spin) Hall conductivity inside the bulk gaps. The Chern number $\mathcal{C}_n$ associated to the band $n$ is an integer topological index defined as the integral of the Berry curvature over the 1BZ,\cite{Kohmoto1985}
\begin{equation}
  \mathcal{C}_n=\frac{1}{2\pi}\int_\mathrm{1BZ}d\vk \left[\partial_{k_x}A^y_{n}(\vk)-\partial_{k_y}A^x_{n}(\vk)\right],
\end{equation}
where $A_{n}^\mu(\vk)=-\ii\langle\psi_{n,\vk}|\partial_{k_\mu}|\psi_{n,\vk}\rangle$ ($\mu=x,y$) is the Berry connection, defined in terms of the eigenstates $|\psi_{n,\vk}\rangle$ of band $n$. We remark that, for this Chern number to be well defined, degeneracies in the spectrum are forbidden. Thus, if two bands touch at a point, the Chern numbers of these bands are ill defined. The problem of the spin degeneracy of the bands in our model is resolved by projecting the system to a single spin component. Thus, we obtain separate Chern numbers for the spin-up and the spin-down bands, which we write as $\mathcal{C}_n^{\spinup}$ and $\mathcal{C}_n^{\spindown}$, respectively.

In the case of the Lieb lattice with only imaginary NNN hopping (i.e., $t'=0$ and $\lambdaISO>0$), we can evaluate the Chern numbers of the bands analytically. Without loss of generality, let us focus on the spin-up component and assume $\lambdaISO>0$. The Hamiltonian can then be written as $\mathcal{H}_{\vk}^\uparrow=\vec{R}\cdot\vec{L}$, where $\vec{R}=(R_1,R_2,R_3)$ with $R_1=-2t\cos(k_x/2)$, $R_2=-2t\cos(k_y/2)$ and $R_3=4\lambdaISO\sin(k_x/2)\sin(k_y/2)$ and $\vec{L}\equiv(L_1,L_2,L_3)$ is the vector of the Gell-Mann matrices
\begin{align}
L_1=\begin{pmatrix}     0 & 1 & 0 \\  1 & 0 & 0 \\  0 & 0 & 0   \end{pmatrix},\ %
L_2=\begin{pmatrix}    0 & 0 & 1 \\  0 & 0 & 0 \\  1 & 0 & 0  \end{pmatrix},\ %
L_3=\begin{pmatrix}    0 & 0 & 0 \\  0 & 0 & -\ii \\  0 & \ii & 0  \end{pmatrix}.\end{align}
These matrices satisfy the commutation relations $[L_i,L_j]=i\epsilon_{ijk}L_k$, so they form a basis for the spin-$1$ representation of the Lie algebra $\mathfrak{su}(2)$. This particular structure of the Hamiltonian allows one to calculate the Chern numbers analytically since here they have a simple interpretation as a winding number. With a computation analogous to the one presented in Ref.~\onlinecite{HeEA2012}, it may be shown that the Berry curvature for the upper and lower band equals
\begin{equation}
\label{eqn_berrycurvature_pm}
  F_\pm
  = \partial_{k_x}A^y_\pm - \partial_{k_y}A^x_\pm
  = \mp\hat{\vec{R}}\cdot(\partial_{k_x}\hat{\vec{R}}\times\partial_{k_y}\hat{\vec{R}}),
\end{equation}
with $\hat{\vec{R}}=\vec{R}/ R$, while for the middle band the Berry curvature vanishes. The integer value of the Chern number then follows from interpretation of the integral of the Berry curvature as the winding number associated to the vector field $\vec{R}$. The result of this computation is $\mathcal{C}_\pm^\uparrow=\mp1$ and $\mathcal{C}_0^\uparrow=0$.\cite{HeEA2012}

In many cases, an analytic computation of the Chern numbers is unfeasible due to the complicated dependence of the eigenvectors on the momentum $\vk$. In such a case, the Chern numbers can be computed numerically from the eigenstates on a discretized lattice of momenta in the 1BZ.\cite{FukuiEA2005} It is shown that in order to obtain the correct values of the Chern numbers, a small number of discretization steps generally suffices. In this paper, we have used the method explained in Ref.~\onlinecite{FukuiEA2005}, and we have confirmed the results with those from analytic computations wherever possible.



%
\end{document}